\documentclass{iaus}
\usepackage{graphicx}

\title[Evolution of magnetic fields in galaxies] 
{Testing the cosmological evolution of magnetic fields in galaxies
with the SKA}

\author[Arshakian et al.]   
{T.G. Arshakian$^1$, R. Beck$^1$, M. Krause$^1$
\and D. Sokoloff$^2$}

\affiliation{$^1$Max-Planck-Institut f\"ur Radioastronomie, Bonn, Germany
              \break email: [tarshakian;rbeck;mkrause]@mpifr-bonn.mpg.de\\[\affilskip]
              $^2$Department of Physics, Moscow State University, Russia
              \break email: sokoloff@dds.srcc.msu.su
             }

\pubyear{2009}
\volume{259}  
\pagerange{119--126}
\date{?? and in revised form ??}
\jname{Proceedings Title IAU Symposium}
\editors{A.C. Editor, B.D. Editor \& C.E. Editor, eds.}
\begin{document}

\maketitle

\begin{abstract}
We investigate the cosmological evolution of large- and small-scale
magnetic fields in galaxies at high redshifts. Results from
simulations of hierarchical structure formation cosmology provide a
tool to develop an evolutionary model of regular magnetic fields
coupled to galaxy formation and evolution. Turbulence in
protogalactic halos generated by thermal virialization can drive an
efficient turbulent dynamo. The mean-field dynamo theory is used to
derive the timescales of amplification and ordering of regular
magnetic fields in disk and dwarf galaxies. For future observations
with the SKA, we predict an anticorrelation at fixed redshift
between galaxy size and the ratio between ordering scale and galaxy
size. Undisturbed dwarf galaxies should host fully coherent fields
at $z<1$, spiral galaxies at $z<0.5$.

\keywords{Techniques: polarimetric,
galaxies: evolution,
galaxies: magnetic fields,
radio continuum: galaxies}
\end{abstract}

\firstsection 
\section{Introduction: importance of magnetic evolution in galaxies}

The observed polarized synchrotron emission and Faraday rotation showed the
presence of regular large-scale magnetic fields with spiral patterns
in the disks of nearby spiral galaxies (Beck 2005), which were
successfully reproduced by mean-field dynamo theory (Beck et al
1996, Shukurov 2005). It is therefore natural to apply dynamo theory
also in predicting the generation of magnetic fields in young
galaxies at high redshifts.

We now have sufficient evidence that strong magnetic fields were
present in the early Universe ($z<3$; Bernet et al. 2008, Seymour et
al. 2008) and that synchrotron emission from distant galaxies
should be detected with future radio telescopes such as the Square
Kilometre Array (SKA). The SKA will allow us
to observe an enormous number of distant galaxies at similar resolution
to that achievable for nearby galaxies today (van der Hulst et al. 2004).
The formation and evolution of regular large-scale magnetic fields
is intimately related to the formation and evolution of disks in
galaxies in terms of geometrical and physical parameters.
A more robust understanding of the history of magnetism in young galaxies
may help to solve fundamental cosmological questions about the
formation and evolution of galaxies (Gaensler et al. 2004).

\section{Three-phase model for the evolution of magnetic fields in galaxies}
\label{sec:greenfun}

We have used the dynamo theory to derive the timescales of
amplification and ordering of magnetic fields in disk and
quasi-spherical galaxies (Arshakian et al. 2008). This has provided
a useful tool in developing a simple evolutionary model of regular
magnetic fields, coupled with models describing the formation and
evolution of galaxies. In the hierarchical structure formation
scenario, we identified three main phases of magnetic-field
evolution in galaxies. In the epoch of \emph{dark matter halo
formation}, seed magnetic fields of $\approx10^{-18}$~G strength
could have been generated in protogalaxies by the Biermann battery or Weibel
instability (first phase). Turbulence in the protogalactic halo
generated by \emph{thermal virialization} could have driven the
turbulent (small-scale) dynamo and amplify the seed field to the
equipartition level of $\approx 20~\mu$G within a few $10^8$ yr
(second phase). In the epoch of \emph{disk formation}, the turbulent
field served as a seed for the mean-field (large-scale) dynamo in
the disk (third phase).

We defined three characteristic timescales for the evolution of
galactic magnetic fields: one for the amplification of the seed
field, a second for the amplification of the large-scale regular
field, and a third for the field ordering on the galactic scale
(Arshakian et al. 2008).
Galaxies similar to the \emph{Milky Way} formed their disk at $z\approx10$.
Regular fields of equipartition (several $\mu$G) strength and a few
kpc coherence length were generated within 2~Gyr (until
$z\approx3$), but field ordering up to the coherence scale of the
galaxy size took another 6~Gyr (until $z\approx0.5$). \emph{Giant galaxies}
had already formed their disk at $z\approx10$, allowing more
efficient dynamo generation of equipartition regular fields (with a
coherence length of about 1~kpc) until $z\approx4$. However, the age
of the Universe is too young for fully coherent fields to have
already developed in giant galaxies larger than about 15~kpc. \emph{Dwarf
galaxies} formed even earlier and should have hosted fully coherent
fields at $z\approx1$.
\emph{Major mergers} excited starbursts with enhanced turbulence, which in
turn amplified the turbulent field, whereas the regular field was
disrupted and required several Gyr to recover. Measurement of
regular fields can serve as a clock for measuring the time since the
last starburst event.
Starbursts due to major mergers enhance the turbulent field strength
by a factor of a few and drive a fast wind outflow, which magnetizes
the intergalactic medium. Observations of the radio emission from
distant starburst galaxies can provide an estimate of the total
magnetic-field strength in the IGM.

This evolutionary scenario can be tested by measurements of
polarized synchrotron emission and Faraday rotation with the SKA. We
\emph{predict}: (i) an anticorrelation at fixed redshift between
galaxy size and the ratio between ordering scale and galaxy size,
(ii) undisturbed dwarf galaxies should host fully coherent large-scale fields at
$z<1$, spiral galaxies at $z<0.5$, (iii) weak regular
fields (small Faraday rotation) in spiral galaxies at $z<3$, but
possibly associated with strong anisotropic fields (strong polarized
emission), would be signatures of major mergers.

\begin{acknowledgments}
This work is supported by the EC Framework Program 6, Square Kilometre Array Design Study (SKADS) and the DFG-RFBR
project under grant 08-02-92881.
\end{acknowledgments}

\end{document}